\documentclass[twocolumn,amsmath,amssymb,aps,prl,10pt,lettersize,superscriptaddress,showpacs]{revtex4-1}

\usepackage[dvips]{graphicx} 
\usepackage{braket}
\usepackage[usenames]{color}
\usepackage{ulem}
\renewcommand{\prl}{Phys.\ Rev.\ Lett.\ }

\newcommand{\PRL}[2]{\prl \textbf{#1}, #2 }

\newcommand{\PNAS}[2]{PNAS \ \textbf{#1}, #2 }
\newcommand{\articletitle}[1]{}

\def\Gakushuin{Faculty of Science, Gakushuin University, 
Mejiro, Tokyo 171-8588,Japan}
\def\UTokyo{Department of Applied Physics, Graduate School of Engineering, The University of Tokyo, Bunkyo-ku, Tokyo 113-8656, Japan}

\begin{document}

\title{Ferromagnetic levitation and harmonic trapping of \\ 
a milligram-scale Yttrium Iron Garnet sphere}

\author{Maria Fuwa}
\email{maria.fuwa.uni@gmail.com}
\affiliation{\Gakushuin}
\date{\today}

\author{Ryosuke Sakagami}
\affiliation{\UTokyo}

\author{Tsuyoshi Tamegai}
\affiliation{\UTokyo}

\begin{abstract} 

We report passive magnetic levitation and three-dimensional harmonic trapping of a 0.3 milligram, 0.5 millimeter diameter Yttrium Iron Garnet sphere at 4 K. 
The gradient of an external magnetic field is used for vertical trapping, while the finite size effect of the diamagnetic effect is used for horizontal trapping. 
The dynamics of the levitated sphere was optically measured to have trapping frequencies of up to around 600 Hz and mechanical $Q$-factors in the order of $Q \sim 10^3$. 
These results were quantitatively reproduced by three-dimensional finite element method simulations. 
Our results can provide a novel system where magnetism, rigid body motions, microwaves, and optics interact. 

\end{abstract}

\maketitle



Levitation of particles in free space provides an experimental platform where external rigid body motion and internal degrees of freedom can freely interact with each other in conditions of extreme environmental isolation. 
The manipulation and measurement of these dynamics with high precision can open way to a wide variety of fundamental and applied research~\cite{KeiselPNAS, TurkerRev, MillenRev, Gonzalez-BallesteroRev, SticklerRev, WinstoneRev}. 
Towards this end, various objects ranging from dielectric nanoparticles~\cite{DelicScience}, nano diamonds~\cite{DelordNature, HuilleryPRB, PerdriatRev}, metals~\cite{RhimRevSci}, hard magnets~\cite{WangPRAp, JiangAPL}, diamagnets~\cite{LewandowskiPRAp, LengPRAp, VinantePRap, RomagnoliArXiv}, to superconductors~\cite{LatorreIEEE, HoferArxiv, LatorrePRAp} have been levitated.

Another promising levitation candidate is single-crystal Yttrium Iron Garnet (YIG) for its large volumetric density of spins of $ \sim 10^{22} \, \mu_{\mathrm{B}} \, \mathrm{cm}^{-3} $ with a long coherence time of $ \sim 1 \, \mu$s~\cite{spinwaves}. 
The quanta of collective spin excitations, or magnons, in YIG crystals have been coherently coupled with both microwaves and optical photons through light-matter interactions~\cite{MaksymovJAP, DanyRev, RameshtiRev, BhoiRev}. 
Utilization of technologies in the field of cavity quantum electrodynamics to enhance interactions~\cite{Haroche-book} has enabled cavity-mediated strong magnon-microwave couplings~\cite{HueblPRL, TabuchiPRL, GoryachevPRAp, ZhangPRL14}, spin current generation~\cite{BaiPRL17}, and superconducting-qubit-magnon coupling~\cite{TabuchiScience} that can resolve single magnons~\cite{DanySciAd} or enable superposition of magnons~\cite{HePRA}. 
Applications of this growing field ranges from microwave-to-optical transducers~\cite{HisatomiPRB,  HaighPRL, ZhangPRL16, GloppePRB}, non-Hermitian physics~\cite{ZhangNatCom}, nonlinear magnonics~\cite{ShenPRL}, magnon lasers~\cite{WangPRA22}, dark matter searches~\cite{TricklePRL, QuaxPRL}, to multimode quantum memories~\cite{ZhangNatC}.

The combination of these magnonic effects with levitation can provide a novel system where magnetism, rigid body motions, microwaves and optics are intertwined. 
First, the collective enhancement in magnon-microwave coupling can compensate for the decrease in coupling of particle position and a control microwave field with increasing mass.
This allows intensive coupling, coupling strength that is nearly independent of the size and mass of the particle, which allows center-of-mass cooling of a levitated subcentimeter-sized particle~\cite{KaniPRLcool}. 
Second, the great reduction in clamping losses by levitation may increase magnetostrictive coupling that can allow ground state cooling of translational motion~\cite{BallesteroPRL, BallesteroPRB}. 
Third, due to the conservation of angular momentum, a change in spin excitation can lead to its mechanical rotation via the Einstein-de-Haas effect~\cite{EdHpaper, WachterJosaB} and its reciprocal Barnett effect~\cite{Barnettpaper}. 
This spin-mechanical coupling may enable transduction of angular momentum in the quantum regime~\cite{RusconiPRB16, RusconiPRL17}, which serves as ultrasensitive torque sensors~\cite{KaniPRL22} or magnetometers~\cite{KimballPRL, FadeevQST}.

\begin{figure}[!b]
\begin{center}
\includegraphics[width= \linewidth, clip]{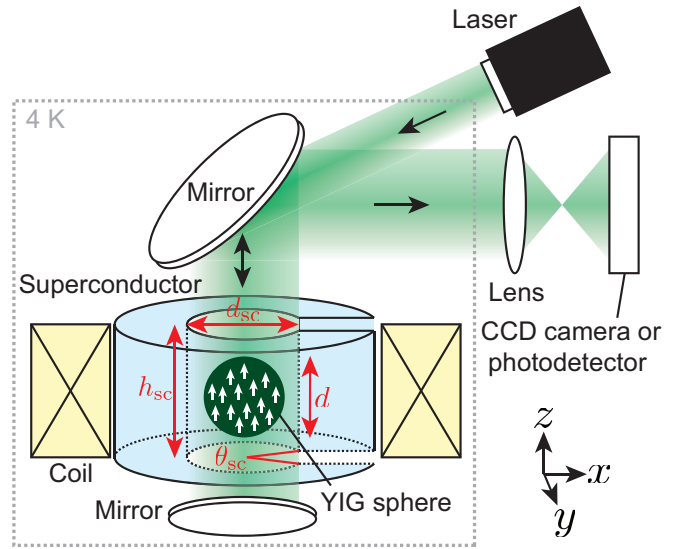}
\caption{Experiment schematic overview.} 
\label{fig:scheme_overview}
\end{center}
\end{figure}

Here we report the experimental demonstration of a magnetically levitated YIG sphere in cryogenic conditions. 
The center of mass motion is three-dimensional harmonically trapped by utilizing the magnetic field gradient of an external magnetic field for vertical trapping, and the finite size effect of the diamagnetic effect from a superconductor whose size is close to the YIG sphere for horizontal trapping~\cite{YIG_lev_theory}. 
Since there is no eddy current damping in a superconductor, our system is a ultimately very low dissipation system. 
We optically read out the motion of the levitated YIG sphere to determine the trap frequency and motional $Q$-factor. 
Finally, the experimental results are compared with three-dimensional finite element method simulations.


Our experimental setup is shown in Fig.~\ref{fig:scheme_overview}. 
A commercial YIG sphere of 0.3 mg mass, $ d = 0.5 $ mm diameter is trapped in a three-dimensional harmonic potential at the center of the hole of a superconductor bulk of inner diameter $ d_{\mathrm{sc}} = 0.76 $ mm, height $ h_{\mathrm{sc}} = 1.04 $ mm, with a slit of angle $ \theta_{\mathrm{sc}} \sim 40^\circ $. 
The YIG sphere diameter of $ d = 0.5 $ mm was chosen for having longest spin coherence time~\cite{TabuchiPRL}. 
A variable external magnetic field is applied to magnetize the Yttrium Iron Garnet (YIG) sphere using a self-made Niobium and Titanium alloy (NbTi) superconducting solenoid coil 
with magnet constant 125 mT/A 
and charging time constant $ 5 \times 10^{-5} $ s. 
To increase the magnet constant, this solenoid coil is designed so that the superconductor bulk barely fits into its bore. 

Since the superconductor bulk in the diamagnetic state focuses this external magnetic field in the vertical direction ($z$ direction), the YIG sphere is trapped in the center of the superconductor hole where the magnetic flux density is maximum. 
The magnetic force and trap frequency are
\begin{gather}
F_z = M V \frac{d B_z}{dz} \\
f_z = \frac{1}{2 \pi} \sqrt{ \frac{M}{\rho} \frac{d^2 B_z}{d z^2}}
\propto \frac{B_z}{d } 
\end{gather}
where $M$ is the magnetization, V volume, $ \rho $ density of the YIG sphere, $B_z$ the $z$ component of magnetic flux density.  

Horizontal ($x, \, y$ direction) stability is achieved by the finite size effect of the diamagnetic effect that arises when dimensions of the hole of the superconductor bulk are comparable to the size of the YIG sphere. 
In this case, the magnetic force on the YIG sphere and resulting trap frequency can only be calculated numerically by integrating the Maxwell stress tensor over its surface.
Since the superconductor bulk is zero-field cooled, 
a slit is introduced to allow the magnetic flux density to enter into the hole. 
The slit shifts the trapping center by a negligible 20 $\mu$m closer to the slit from the center of the hole. 

We compare our levitation scheme with two kinds of superconductors. 
One is a cubic melt processed YBCO with the $c$ axis, the direction perpendicular to the planes of copper-oxide layers, aligned with the direction of external magnetic field ($z$ direction in Fig.~\ref{fig:scheme_overview}). 
This has a critical temperature $ T_\mathrm{c} \sim 93 $ K~\cite{firstYBCO}, 
and magnetic field tolerance $ H_{\mathrm{pen}} (0 \, \mathrm{K}) \sim $ 1.2 T (Appendix B). 
Here $ H_{\mathrm{pen}} $ is the maximum magnetic field our superconductor sample could expel by perfect cancellation of their magnetic moments, which depends not only on the type of superconductor, purity, and structural properties but also highly on shape. 
Since both the transition temperature and critical magnetic fields are high, YBCO is expected to retain its superconductivity for higher magnetic fields, enabling higher trapping frequency limits up to 1 kHz for a sphere of the same size, and over 10 kHz for spheres below 100 $\mu$m in diameter~\cite{YIG_lev_theory}.   
This would facilitate ground state cooling of the center-of-mass motion of micro-gram-scale levitated YIG spheres in the future. 

The other is a high purity single grain Niobium (Nb) with a residual-resistance ratio RRR = 582, $ T_{\mathrm{c}} \sim 9.3 $ K, 
and $ H_{\mathrm{pen}} (0 \, \mathrm{K}) \sim $ 125 mT (Appendix B).  
The high RRR indicates that the superconductor is relatively pure with fewer impurities that can disrupt the flow of electrical current through the material. 
This leads to better superconducting properties and less dissipation from residual resistance which can lead to eddy current damping. 
Furthermore, it is more machinable compared to high temperature superconductors, which can lead to increased design flexibility, improved surface finish and lower manufacturing costs with faster production times. 

The trapping system is cooled to 4 K using a pulse-tube cryocooler (PT407, Cryomech, Inc.). 
This cryostat is pumped to a pressure of $P = 2 \times 10^{-4} $ Pa with a rotary vane pump and turbo pump. 
Then a 5 mW laser at 532 nm (Verdi, Coherent, Inc.) is injected into the cryostat from a window to illuminate the YIG sphere and superconductor. 
Some of the light is reflected to pass through a telescope with a long focus length of $ f = $ 250 mm to measure the movement of the YIG sphere using either a CCD camera (DCC1645C-HQ, Thorlabs, Inc.) or photo-detector (PDA36A-ED, Thorlabs, Inc.). 
To reduce seismic noise, the cryostat was vacuum-sealed, and both the cryocooler and vacuum pumps were turned off during data acquisition.


\begin{figure}[!t]
\begin{center}
\includegraphics[width= 5 cm, clip]{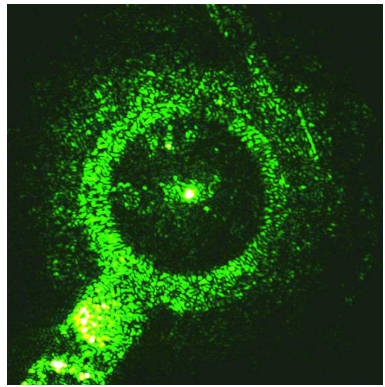}
\caption{Photo of a levitated $ d = 0.5 $ mm YIG sphere trapped in the center of the YBCO hole observed using a CCD camera.} 
\label{fig:lev_photo}
\end{center}
\end{figure}

Initially the YIG sphere is placed at the bottom of the superconductor hole.
Then it is lifted up by applying an external magnetic field above $ B_z \geq $ 7.5 mT. 
This excites a vertical oscillation, as the YIG sphere is trapped at the center of the hole (Fig.~\ref{fig:lev_photo}). 
To measure the oscillations, a photo-detector is placed at the end of the telescope to monitor the change in optical intensity as the YIG sphere moves. 
1,250,000 points were acquired using a data-logger (DLM2025, Yokogawa Electric Corporation) at a sampling frequency of 125 kHz for a measurement time of 10 seconds. 
The base plate temperature was monitored during this measurement, which started at 3.8 K and increased to a maximum of 4.6 K.

\begin{figure}[!t]
\begin{center}
\includegraphics[width= 7.5 cm, clip]{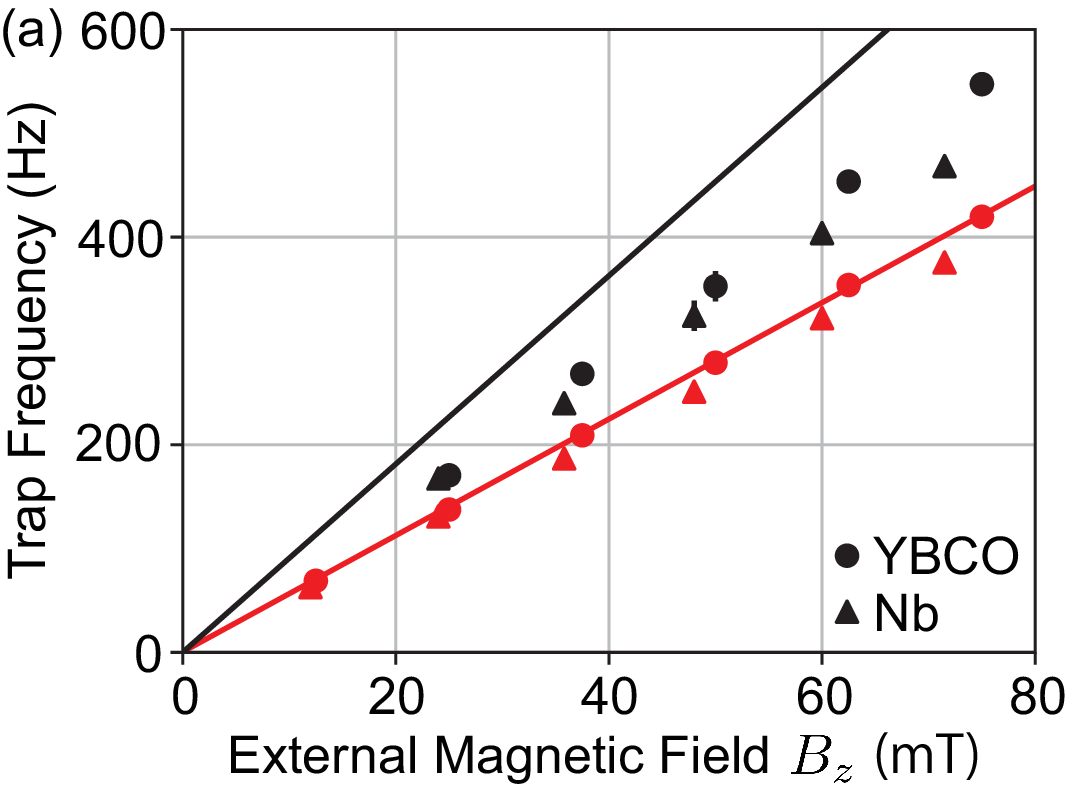}
\vspace{3mm}
\includegraphics[width= \linewidth, clip]{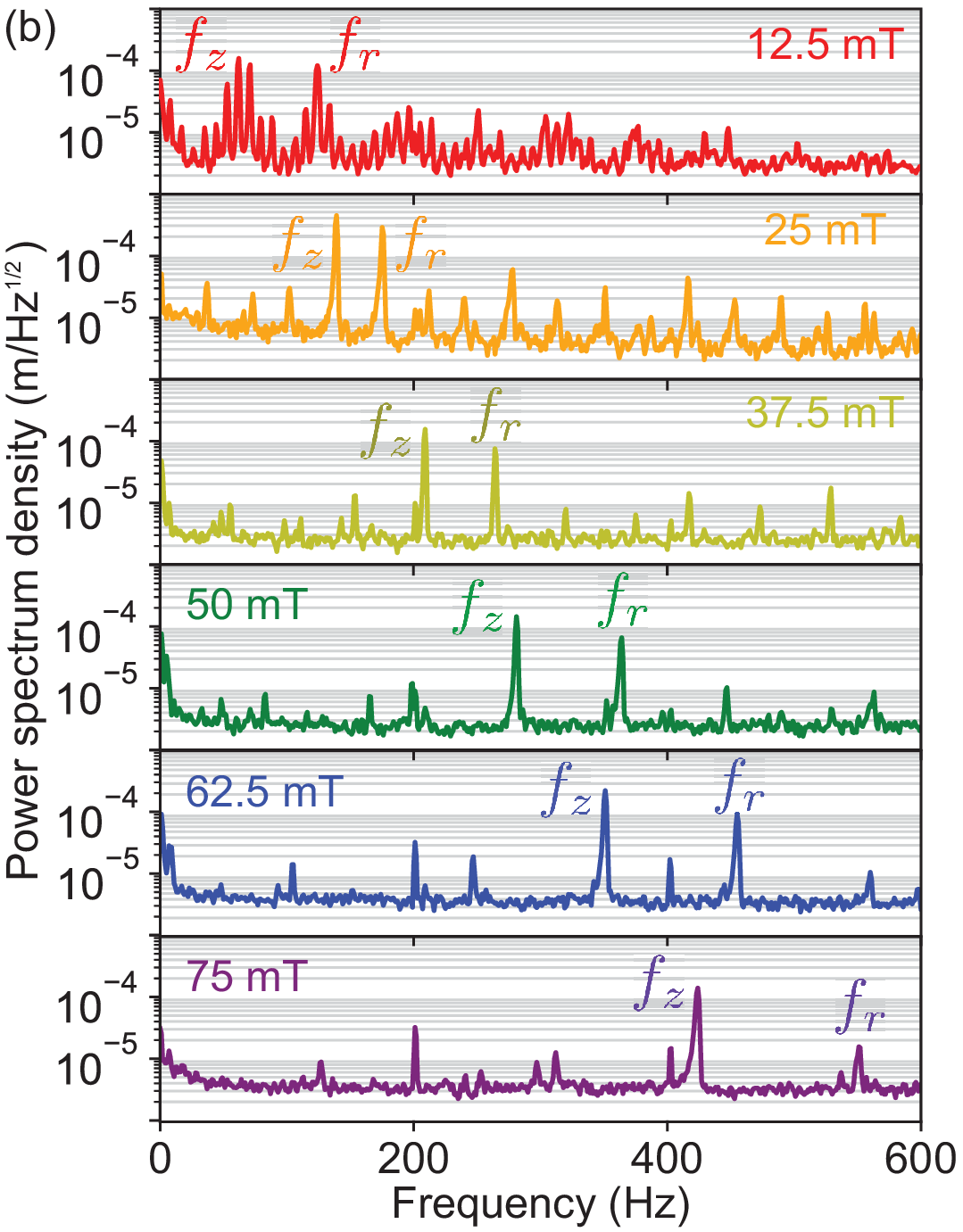}
\caption{(a) Trap frequency for varying external magnetic fields $B_z$. 
Red: vertical oscillation $f_z$ ($z$ direction in Fig.~\ref{fig:scheme_overview}), black: horizontal oscillation $f_y$ ($y$ direction in Fig.~\ref{fig:scheme_overview}), solid line: simulated trap frequencies, plotted with experimentally observed frequencies for circle: YBCO, triangle: Nb as superconductor respectively.
(b) Power spectrum density for varying external magnetic fields $B_z$. $f_z$: vertical oscillations, $f_r$: diagonal oscillations.} 
\label{fig:result_freq}
\end{center}
\end{figure}

Modes have been identified by comparing the trapping frequencies to the frequencies predicted for this configuration by a three-dimensional finite element analysis using COMSOL Multiphysics (Appendix C). 
Since the easy axis of the levitated sphere is aligned with $B_z$, it is rotationally symmetric, and only the translation oscillations are detected. 
Figure~\ref{fig:result_freq} (a) shows the dependence of the trapping frequency on the external magnetic field density $B_z = 12.5 - 75$ mT for superconductors YBCO and Nb. 
Since $B_z$ is below the saturation magnetization of YIG at $ M_s = 196 $ kA/m, the trapping frequencies should linearly increase with $B_z $ as 
\begin{gather}
f_z \, [\mathrm{Hz}] = 5.6 \times 10^{-3} \, B_z  \, [\mathrm{mT}] \\
f_y \sim 1.6 f_z, \quad f_x \sim 1.5 f_z
\end{gather}
This is variable in the range of 50 Hz to as high as 600 Hz. 
The maximum trapping frequency is limited by the maximum current of 0.6 A that can be applied to our home-made superconducting coil due to joule heating. 

For YBCO, the vertical trapping frequencies matches the simulated prediction.
Since the initial position of the YIG sphere is slightly shifted from the center of the superconductor hole, diagonal oscillations are also excited during the lift-off process. 
These diagonal oscillations have frequencies between $ f_z \leq f_r \leq f_y $ which match the theoretical prediction.

In the case of Nb, both the vertical and diagonal trapping frequencies are close to the simulated prediction. 
However, the discrepancy increases with the external magnetic field. 
This is likely due to the fact that the system temperature of 4 K is close to the critical temperature $T_c$ of Nb, and restoring force from the diamagnetic effect weakens as the persistent current approaches the critical current $J_c$. 
Thus it is likely that Nb has an effective inner diameter larger than the actual diameter $ d_{\mathrm{sc}} $, leading to a slightly weaker trapping potential and lower trapping frequency. 

A typical frequency spectrum of the levitating YIG sphere shows how the magnetic trapping gets stronger with increasing $B_z$ (Figure~\ref{fig:result_freq} (b)). 
The amplitude of the oscillations have been calibrated from the maximum oscillation amplitude being the lift distance of 0.25 mm. 
For $ B_z = $ 12.5 mT, the magnetic spring is weak and multiple peaks are observed near the expected trapping frequency, indicating a fluctuation in YIG sphere position. 
Since the vertical and horizontal trapping frequencies are close, it is difficult to resolve these modes.
As the external magnetic field increases to $B_z \geq $ 37.5 mT, distinct vertical and diagonal oscillations can be observed, indicating that only the desired modes are excited.

\begin{figure}[!t]
\begin{center}
\includegraphics[width= 7.5 cm, clip]{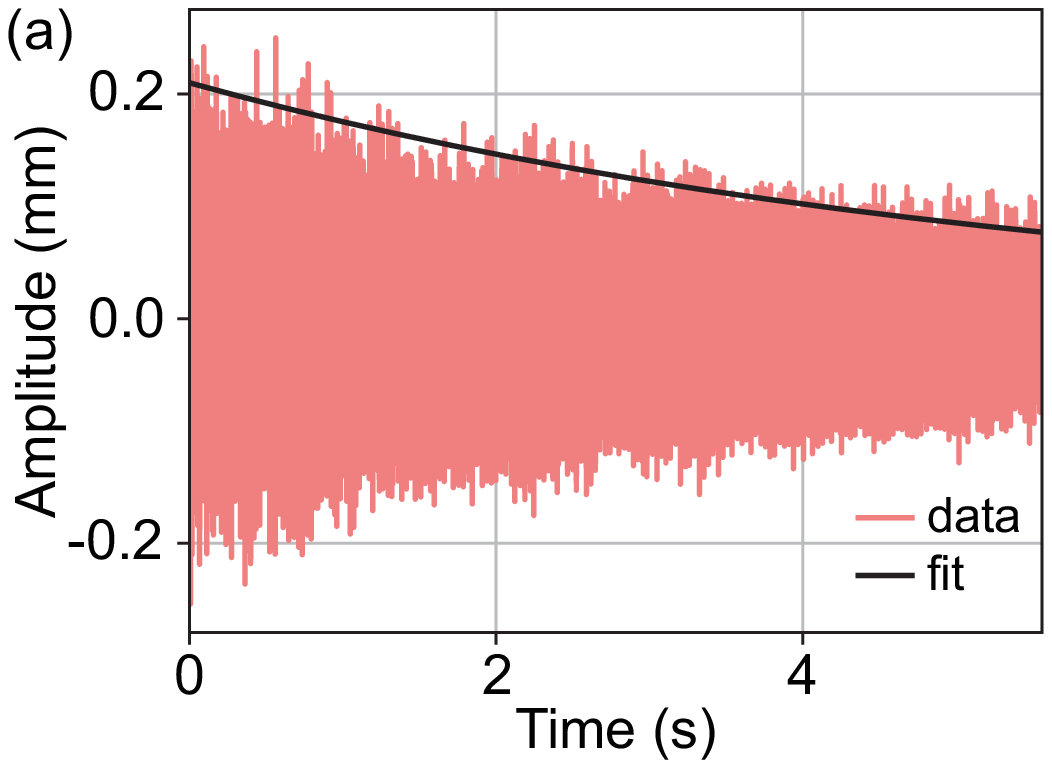}
\vspace{3mm}
\includegraphics[width= \linewidth, clip]{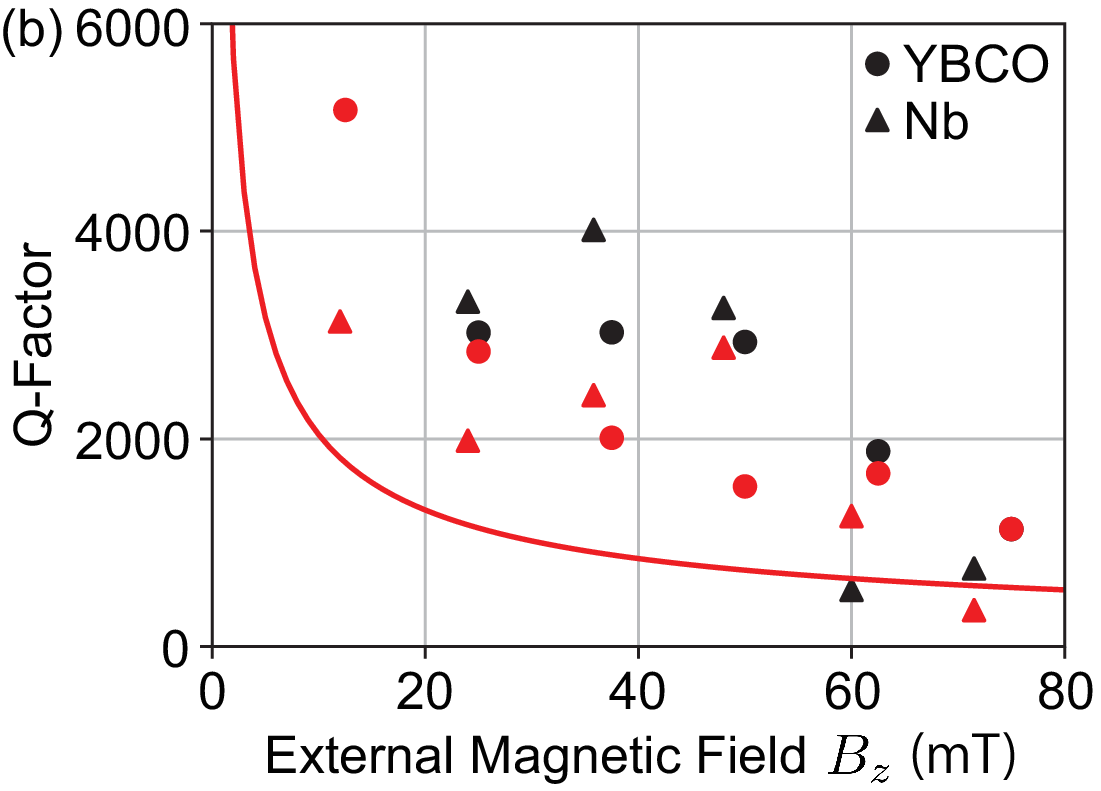}
\caption{(a) Example of measured amplitude decay for $ B_z = 25 $ mT, $ f_r = 169 $ Hz. 
(b) Mechanical $Q$-factors for varying external magnetic fields $B_z$. 
Circle: YBCO as superconductor, triangle: Nb as superconductor, red: vertical oscillations at frequency $f_z$, black: diagonal oscillations at frequency $f_r$ in Fig.~\ref{fig:result_freq}. 
Red line: theoretically predicted $Q$-factor for the vertical oscillation mode ($z$ direction in Fig.~\ref{fig:scheme_overview}) at frequency $f_z$.} 
\label{fig:result_Qfactor}
\end{center}
\end{figure}


The $Q$-factor is measured by exciting oscillations by lifting up the YIG sphere, and observing the amplitude decay of the envelope.  
The frequency modes were isolated with the use of a band pass filter, and the decay is fitted to an exponential function of time constant $\tau$, with a relative fitting error below $ < 10^{-3}$.  
Finally, the $Q$-factor is calculated as $ Q = 2 \pi f_i / \tau $, where $f_i$ is the trap frequency obtained by Fourier transforming the decaying oscillations. 
The resolution of the trapping frequency $f_i$ is limited to 0.1 Hz due to the measurement time of 10 s. 

All modes had $Q$-factors in the order of $10^3$, which decreased with increasing $B_z$ (Fig.~\ref{fig:result_Qfactor}). 
No significant $Q$-factor difference was observed between YBCO and Nb as superconductor. 
The $Q$-factor limit from eddy current damping was predicted by a three-dimensional finite element analysis using COMSOL Multiphysics (Appendix C), assuming the electrical conductivity $ \sigma = 6.58 \times 10^9 $ S/m for pure oxygen-free copper at 10 K~\cite{EkinBook}.
This scales as $ Q_{\mathrm{eddy}, \, z} \propto B^{-0.5} $ and is inversely proportional to the electric conductivity $ Q_{\mathrm{eddy}, \, z} \propto \sigma ^{-1} $. 
The experimentally measured $Q$-factors agreed in order with simulated predictions, and obeyed the same tendency of decreasing with external magnetic field. 
The discrepancy is likely due to the actual electrical conductivity of oxygen free copper being lower than $\sigma$, due to material variability or local heating by the coil currents and injected light, and the Q factor limited by the eddy current damping could be higher. 
These results suggest that eddy currents is likely to be the dominant loss in our system. 
Experimental optimization of external magnetic field $B_z$ will be conducted elsewhere. 

The $Q$-factors for horizontal oscillations ($x, \, y$ direction in Fig.~\ref{fig:scheme_overview}) were expected to have an order higher Q factor compared to vertical oscillations ($z$ direction in Fig.~\ref{fig:scheme_overview}). 
This is likely because vertical oscillations induce eddy currents that circulate in a helical pattern around the cylinder's axis, which create a secondary magnetic field aligned with the YIG magnetization, and therefore lead to significant damping. 
On the other hand, horizontal oscillations induce currents around the circumference of the cylinder in a direction perpendicular to its axis, which can create a secondary magnetic field perpendicular with the YIG magnetization, and therefore lead to less damping. 
Since the diagonal oscillations are a combination of vertical and horizontal oscillations, they have higher $Q$-factors than vertical oscillations.


These eddy current damping can be significantly reduced to $ Q_{\mathrm{eddy}, \, z} > 10^8 $, by 
either increasing the bore diameter of the superconducting coil to over 40 mm, using a Helmholtz coil configuration, adding slits to the bobbin to mitigate eddy current flow, and positioning any metal planer stages over 10 mm away. 
In addition, utilization of dielectric material such as fused silica or sapphire for jigs instead of oxygen free copper is expected to significantly reduce the eddy current damping. 
Meanwhile, the gas damping limited $Q$-factor is expected to be over $ Q_{\mathrm{gas}} > 10^{11} $ for our system.
Thus, our system has the potential to reach eddy current limited $Q$-factors over $10^8$ for spheres below 0.2 mm in diameter~\cite{YIG_lev_theory}.  

This high $Q$-factor together with intensive spin-mechanical coupling can enable phononic and magnonic ground state cooled YIG spheres below 0.15 mm in diameter with a trapping  frequency above $f_z > 10$ kHz, under an external magnetic field of $B_z = 1.2$ T, ferromagnetic resonance of 30 GHz and linewidth below 0.3 MHz at 10 mK~\cite{KaniPRLcool, YIG_lev_theory}. 
Applications range from magnonic quantum networks~\cite{RusconiPRA}, quantum tunneling~\cite{WernsdorferSci}, to generation of spatial superpositions~\cite{NairArxiv}. 
Furthermore, since a levitated YIG sphere can be an isolated quantum system which conserves the angular momentum, it may allow coherent superpositions of quantized rotational states through spin mechanical interactions or gyromagnetic effects, which enable highly sensitive gyroscopes~\cite{ZhangOptExp}.
These rotational states may be read out using optical birefringence or by attaching metasurface polarizers~\cite{KurosawaOptExp} for polarization extinction measurements. 


In conclusion, a YIG sphere of 0.3 mg mass, 0.5 mm diameter is levitated and three-dimensional harmonically trapped in the center of a superconductor hole with slit. 
The trap frequency increased linearly with the external magnetic field strength, and was variable between 50 Hz to up to 600 Hz. 
The $Q$-factor for translational oscillations was limited in the order of $Q \sim 10^3$ due to eddy current damping. 
A three-dimensional finite element method is used to explain the levitation dynamics and main source of dissipation of the particle’s motion, of which qualitatively reproduced the experimental results. 
This research lays the foundation for future research and development of levitated YIG systems that can interact with microwaves, superconducting qubits, acoustic phonons, or optical photons.


\section{Acknowledgements}
The authors acknowledge Keita Takahashi for supplying the YBCO bulk as well as valuable discussions on sample preparation, the comments and suggestions from Jason Twamley, Kani Mohamed, Yasunobu Nakamura, Koji Usami, Rekishu Yamazaki, Kohei Matsuura as well as Nobuyuki Matsumoto for fruitful discussions on oscillator characterization. 
The ultra high purity niobium was provided by ULVAC, Inc. KEK Mechanical Engineering Center fabricated the niobium specimen.
This work was supported by JST, PRESTO (grant number JPMJPR1866) and Grant for Basic Science Research Projects from The Sumitomo Foundation (grant number 210825), Shimadzu Science Foundation (grand number 220046), The Mazda Foundation (grant number 22KK-114), Research Foundation for Opto-Science and Technology, and Yazaki Memorial Foundation for Science and Technology.

\section{Appendix}
\section{Appendix A: Experimental setup details}

The YIG sphere is a pure single crystal grown using the floating zone method, diced, and polished into spheres by Microsphere, Inc.
The YBCO is cubic melt processed commercially made by Nippon Steel Corporation and machined into a Landolt ring shape, with the $c$ axis, the direction perpendicular to the planes of copper-oxide layers, aligned with the axis of the cylindrical hole. 
The high RRR Niobium (Nb) ingot is custom grown using electron beam melting by ULVAC, Inc. and machined into a Landolt ring shape by KEK Mechanical Engineering Center. 

The Niobium and Titanium alloy (NbTi) superconducting solenoid coil has a winding inner diameter of 3.1 mm, outer diameter 13 mm, height 1 mm, total number of turns 475 using a wire of diameter 0.1 mm, inductance 1.3 mH and charging time constant $ 5 \times 10^{-5} $ s. 
It is designed with a bore diameter $d_{\mathrm{bore}} = 2.274$ mm that is slightly larger than the superconductor bulk to enlarge the magnet constant of 125 mT/A. 

\begin{figure}[!t]
\begin{center}
\includegraphics[height=4.5cm, clip]{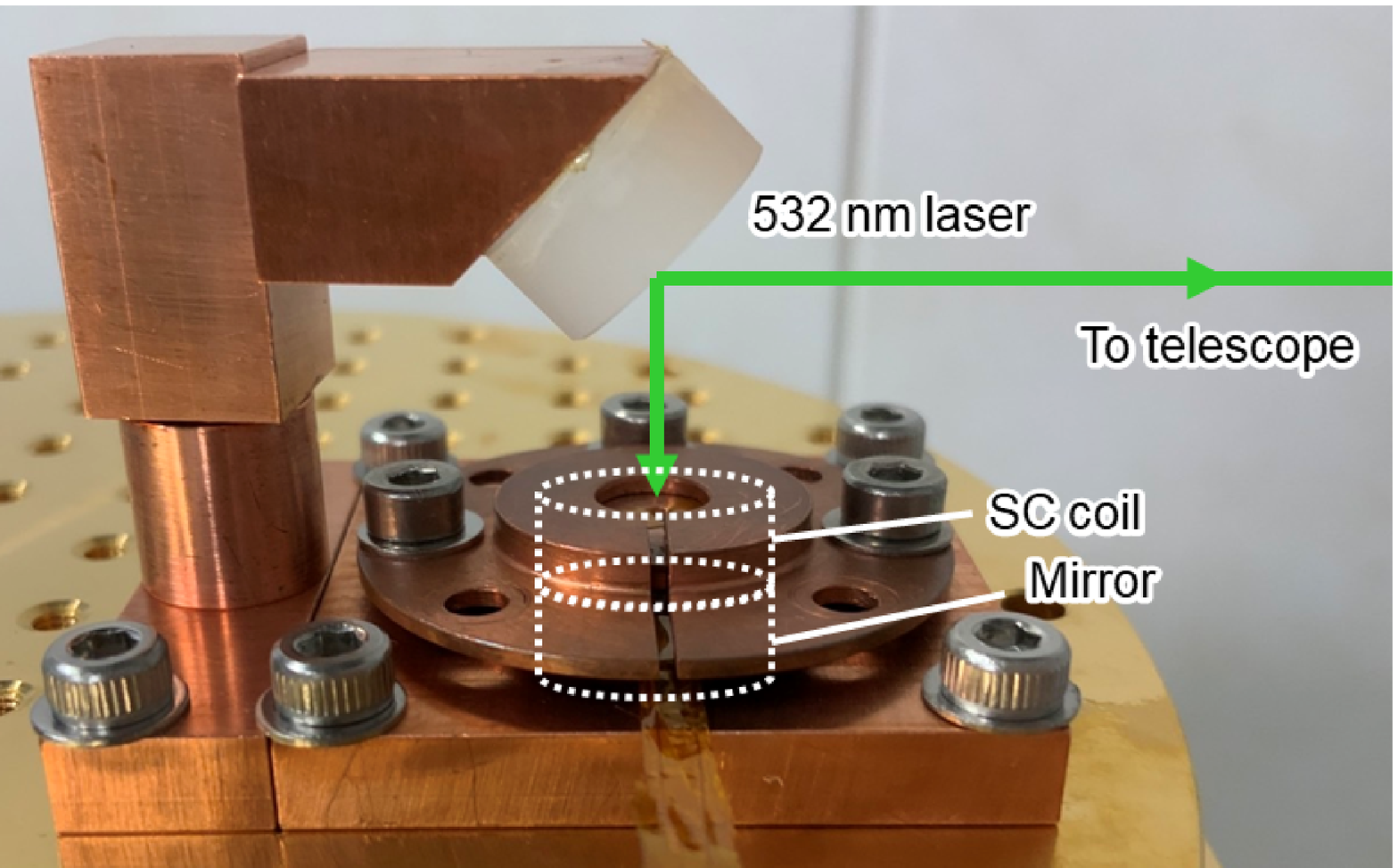}
\vspace{3mm}
\includegraphics[height=4.5cm, clip]{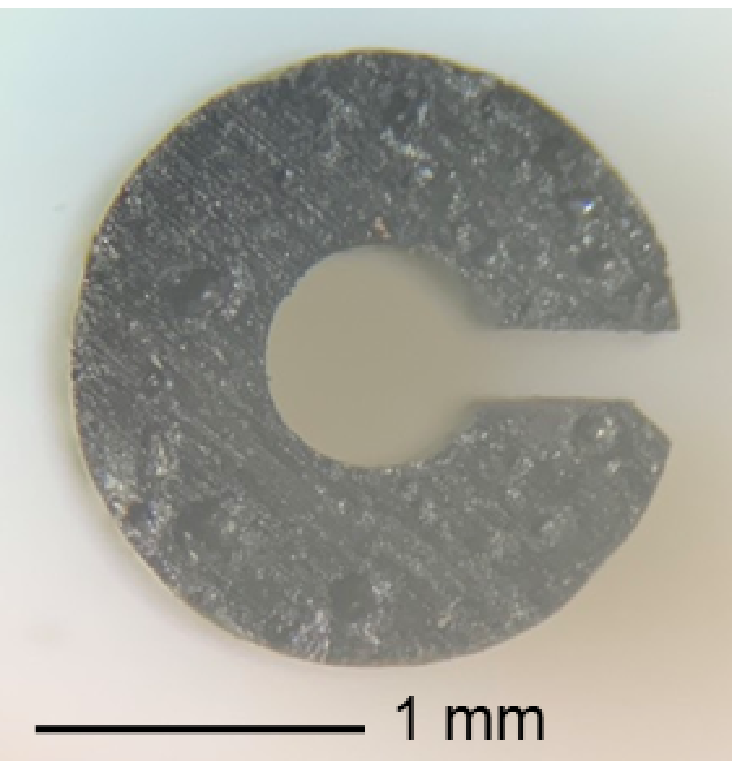}
\vspace{3mm}
\includegraphics[height=4.5cm, clip]{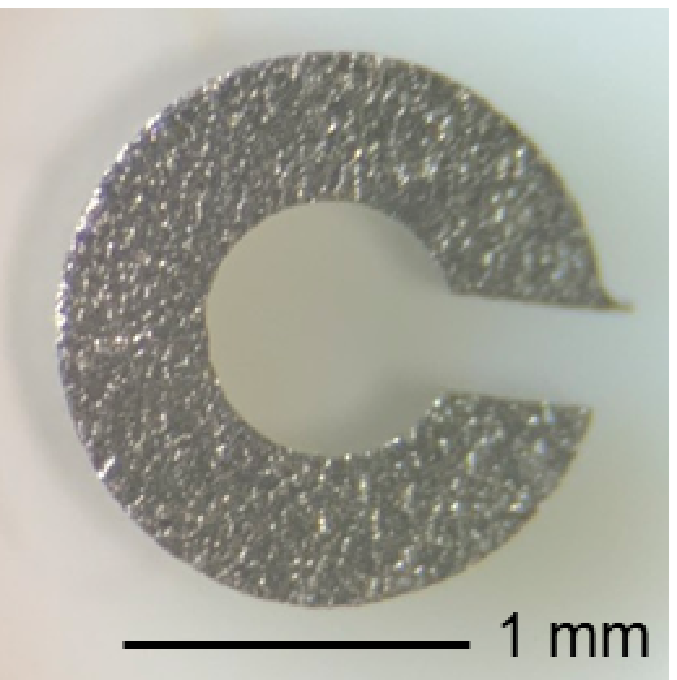}
\caption{(Top) Photograph of experimental setup, 
(middle) YBCO sample, 
(bottom) Niobium sample.} 
\label{fig:exp_photo}
\end{center}
\end{figure}

\begin{table*}[!t]
\centering
\caption{Summary of experimental parameters}
\begin{tabular}{lccc}
\hline
Quantity & Symbol & Value & Unit \\ \hline \hline  
YIG sphere diameter & $d$ & 0.5 & mm \\
YBCO bulk hole inner diameter & $ d_{\mathrm{sc}} $ & 0.756 & mm \\
YBCO bulk outer diameter & $ d_{\mathrm{sc} \, \mathrm{out}} $ & 1.948 & mm \\
YBCO bulk height & $ h_{\mathrm{sc}} $ & 1.039 & mm \\
YBCO bulk slit angle & $ \theta_{\mathrm{sc}} $ & 39 & degrees \\
Nb bulk hole inner diameter & $ d_{\mathrm{sc}} $ & 0.758 & mm \\
Nb bulk outer diameter & $ d_{\mathrm{sc} \, \mathrm{out}} $ & 1.684 & mm \\
Nb bulk height & $ h_{\mathrm{sc}} $ & 1.029 & mm \\
Nb bulk slit angle & $ \theta_{\mathrm{sc}} $ & 42 & degrees \\
Superconductor coil wire diameter & $d_{\mathrm{coil}} $ & 102 & $\mu$m \\
Superconductor coil number of turns & $N_{\mathrm{coil}}$ & 475 &  \\
Superconductor coil height & $h_{\mathrm{mag}}$ & 1 & mm \\
Bobbin bore diameter & $d_{\mathrm{bore}}$ & 2.274 & mm \\
Bobbin thickness &  & 0.5 & mm \\
Bobbin outer diameter &  & 14 & mm \\
Bobbin height &  & 2 & mm \\
Copper lid inner diameter &  & 6 & mm \\
Copper lid outer diameter &  & 14 & mm \\
Copper lid thickness &  & 1 & mm \\ \hline 
YIG density &  & 5172 & $\mathrm{kg/m^3}$ \\
YIG sphere mass & $m$ & 0.339 & mg \\
YIG relative permeability~\cite{YIGmu-init} & $\mu_\gamma$ & 32 &  \\ 
YIG relative permittivity~\cite{YIGdielectric} & $\varepsilon_\gamma$ & 15 &  \\ 
Conductivity of oxygen free copper & $\sigma_{\mathrm{Cu}} $ & $ 6.58 \times 10^9 $ & S/m \\ 
Conductivity of insulators &  & 1 & S/m \\ 
Air and copper relative permeability &  & 1 &  \\ \hline
\end{tabular}
\label{tbl_exp_param}
\end{table*}

The YIG sphere is enclosed inside the superconductor hole, by attaching a piece of microscope cover-glass of thickness $\sim$ 0.17 mm on the top and bottom of the superconductor bulk hole as lids with varnish (GE7031). 
This superconductor piece is then glued onto the superconductor coil bore with varnish to ensure enough heat conduction. 
Thus the initial YIG position is at the bottom of the superconductor bulk hole, and is lift up and trapped in the center when an external magnetic field is applied. 
This structure is placed beneath a half-inch mirror and positioned on a 4 K stage (Fig.~\ref{fig:exp_photo}). 
Note that at the center of the superconductor hole, the external magnetic field $B_z$ is focused in the vertical direction, and is minimum in the horizontal direction. 
If the superconductor does not exhibit diamagnetism, the YIG sphere will stick to the side of the superconductor hole where the horizontal magnetic field is maximum, and therefore no oscillations will be observed. 
A summary of experimental parameters are given in Table~\ref{tbl_exp_param}.

\subsection{Optical heating}

Visible light shown onto the YIG sphere for illumination, can lead to local temperature increase of the levitated YIG sphere. 
Here a lens of focus length $f = 750$ mm is used to focus the optical beam to a diameter of $ d_{\mathrm{opt}} \sim 0.9 $ mm at the YIG sphere. 
Out of this $P_{\mathrm{opt}} = 5$ mW input beam, 
\begin{equation} 
\frac{ P_{\mathrm{opt}} }{ \sqrt{2 \pi} d_{\mathrm{opt}} / 2 } \int_{-d/2}^{d/2} \exp \left( - \frac{x^2}{d_{\mathrm{opt}}} \right) dx 
\sim 2.8 \, \mathrm{mW} 
\tag{A1} 
\end{equation} 
hits the $d=0.5$ mm diameter YIG sphere. 
For a 532 nm laser, with reflectivity of $R_{\mathrm{532 \, nm}} \sim 0.17$~\cite{YIGrefl} for a polished bulk YIG, and absorption coefficient $ \alpha = 10^3 \, \mathrm{cm}^{-1} $~\cite{YIGabsorb}, 
$\dot{q}_{\mathrm{opt}} = 2.35$ mW of the illuminated light is absorbed. 
During the measurement time of 10 s, the YIG sphere is optically heated by 
$\Delta E_{\mathrm{opt}} = 23.5$ mJ.

Since the cooling power of our cryostat of 0.5 W for the superconductor bulk and YIG sphere is much larger than $\dot{q}_{\mathrm{opt}}$, we assume only the temperature of the levitated YIG sphere increases to $T_2$, and everything nearby remains at $ T_1 = 4$ K. 
Bulk YIG has a thermal conductivity of $\kappa_{\mathrm{YIG}} (\mathrm{4 \, K}) = 10.5 $ W/m K and specific heat of $C_{\mathrm{v}} (\mathrm{4 \, K}) = 0.02 $ J/kg K, both of which increase with temperature~\cite{YIGCv}. 
The thermal conductivity throughout the YIG sphere is approximately, 
\begin{equation}
\kappa_{\mathrm{YIG}} \times \pi \left( \frac{d}{2} \right)^2 \times \frac{1}{d} \times ( T_2 - T_1 ). 
\tag{A2} 
\end{equation} 
This gives 8.27 mW for $T_2 = 5$ K, which is larger than $ \dot{q}_{\mathrm{opt}}$. 
Thus we can assume the temperature is uniform over the sphere. 

The heat capacity of the YIG sphere can be modeled using the Debye model as 
\begin{equation}
C_{\mathrm{v}} (T) \propto \left( \frac{T}{T_{\mathrm{D}}} \right)^3 \int_0^{\frac{T_{\mathrm{D}} }{T}} \frac{x^4 e^x}{ (e^x -1)^2 } \, dx 
\tag{A3} 
\end{equation}
with Debye temperature $T_{\mathrm{D}} = 531$ K, and proportional coefficient 2085 J/kg K~\cite{YIGCv}. 
The optical heating can increase the YIG sphere temperature up to $T_2 \sim 246$ K, which satisfies 
\begin{equation} 
\Delta E_{\mathrm{opt}} 
\sim m \int_{T_1}^{T_2} C_{\mathrm{v}} ( \tau ) \, d \tau 
= 23.7 \, \mathrm{mJ}, 
\tag{A4} 
\end{equation} 
with $C_{\mathrm{v}} (246 \, \mathrm{K}) = 556.6 $ J/kg K. 
This is due to the large absorption of visible light by YIG, as well as the small mass $m = 0.339$ mg of the YIG sphere. 

In this case, the radiative heat transfer from the levitated sphere to the glass plates on top and bottom of the superconductor bulk hole is negligible 
\begin{equation}
\dot{q}_{\mathrm{rad}} = \sigma_{\mathrm{SB}} E A_{\mathrm{YIG}} ( T_2^4 - T_1^4 )
\sim 8.6 \, \mu\mathrm{W}. 
\tag{A5} 
\end{equation}
Here 
$\sigma_{\mathrm{SB}} = 5.67 \times 10^{-8} \, \mathrm{W/m^2 \, K^4}$ is the Stefan-Boltzman constant, 
\begin{equation*}
E = \frac{ \varepsilon_1 \varepsilon_2 }{ \varepsilon_2 + A_{\mathrm{YIG}} / A_{\mathrm{env}} ( \varepsilon_1 - \varepsilon_1 \varepsilon_2 )}
\end{equation*}
with emmitivity $ \varepsilon_1 = \varepsilon_2 = 0.9 $ for non-metalic surfaces,  $A_{\mathrm{YIG}} = 7.85 \times 10^{-7} \, \mathrm{m}^2 $ the surface area of the YIG sphere, and $ A_{\mathrm{env}} = 1.3 \times 10^{-4} \, \mathrm{m}^2 $ the surface area of the two glass plates~\cite{EkinBook}. 
Since the emmitivity scales as the square root of a material's resistivity for metals, there is no heat transfer to the superconductor bulk, which reflects all radiative heat. 

The heat conduction through gas in the free molecular regime is a negligible 
\begin{equation}
\dot{q}_{\mathrm{gas}} 
= 1.2 \times \frac{0.5}{1 + 0.5 A_{\mathrm{env}} / A_{\mathrm{YIG}} } P A_{\mathrm{YIG}} ( T_2 - T_1 ) 
\sim 0.45 \, \mu\mathrm{W}, 
\tag{A6} 
\end{equation}
for $P = 2 \times 10^{-3}$ Pa~\cite{EkinBook}. 

Thus, it is possible that the temperature of the levitated YIG increased to as high as $T_2 \sim 246$ K. 
During the experiment, the YIG sphere was repetitively cooled back down to $T_2 = T_1 = 4$ K in between measurements, when the YIG sphere is lowered back to its initial position at the bottom of the superconductor hole on top of a glass plate. 
For quantum technology applications, optical heating can be circumvented by attaching a metasurface mirror to the YIG sphere, or using an infrared laser near 1.55 $\mu$m, where YIG is transparent.

\section{Appendix B: Superconductor properties}

Although the superconducting properties of bulk YBCO and Nb are well known, the properties of the actual material used, the effect of shape and damage applied through machining is unknown.
To asses this, we measured the superconducting properties of samples machined from the same bulk to estimate the properties of the samples in Fig.~\ref{fig:exp_photo}.

\subsection{Critical temperature $T_{\mathrm{c}}$ measurement}

The superconductor critical temperature $T_{\mathrm{c}}$ can be determined by measuring the magnetic moment of the superconductor as a function of temperature, while the material is subjected to a low external magnetic field of 0.5 mT (5 Oe). 
Below $T_{\mathrm{c}}$, the superconductor will expel the magnetic field due to diamagnetic shielding effect, leading to a drop in the magnetic moment. 
The magnetic moment was measured using a superconducting quantum interference device (SQUID) magnetometer (MPMS-XL5, Quantum Design, Inc.). 
The critical temperature is determined as the temperature at which the magnetic moment begins to decrease rapidly.

\begin{figure}[!t]
\begin{center}
\includegraphics[height=5cm, clip]{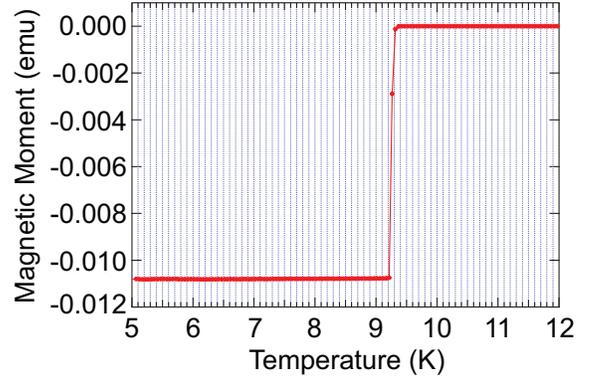}
\caption{Magnetic moment temperature dependence for Nb.} 
\label{fig:MT_meas}
\end{center}
\end{figure}

We measured the critical temperature for the custom made high RRR Nb sample, which had a transition temperature of $T_{\mathrm{c}} \sim 9.3$ K (Fig.~\ref{fig:MT_meas}).
This agrees with the bulk value of $T_{\mathrm{c}} = 9.26 $ K for Nb~\cite{SupercondRev}. 
The sharp drop in the magnetic moment of Nb reflects its purity and good superconducting properties. 
Note that since the YBCO ring is made from a commercial bulk, and its transition temperature of $T_{\mathrm{c}} = 93$ K~\cite{firstYBCO} is well above our levitation conditions at 4 K, indicating that it can be assumed to be in a superconducting state albeit temperature fluctuations, the transition temperature of a machined sample has not been measured.

\begin{figure}[!t]
\begin{center}
\includegraphics[height=5cm, clip]{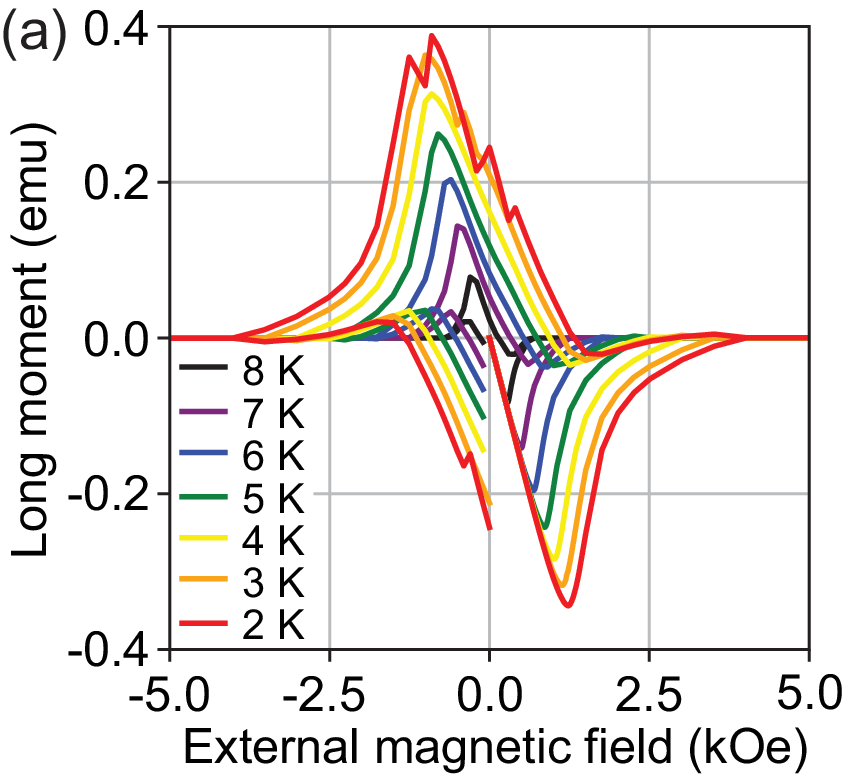}
\vspace{3mm}
\includegraphics[height=10cm, clip]{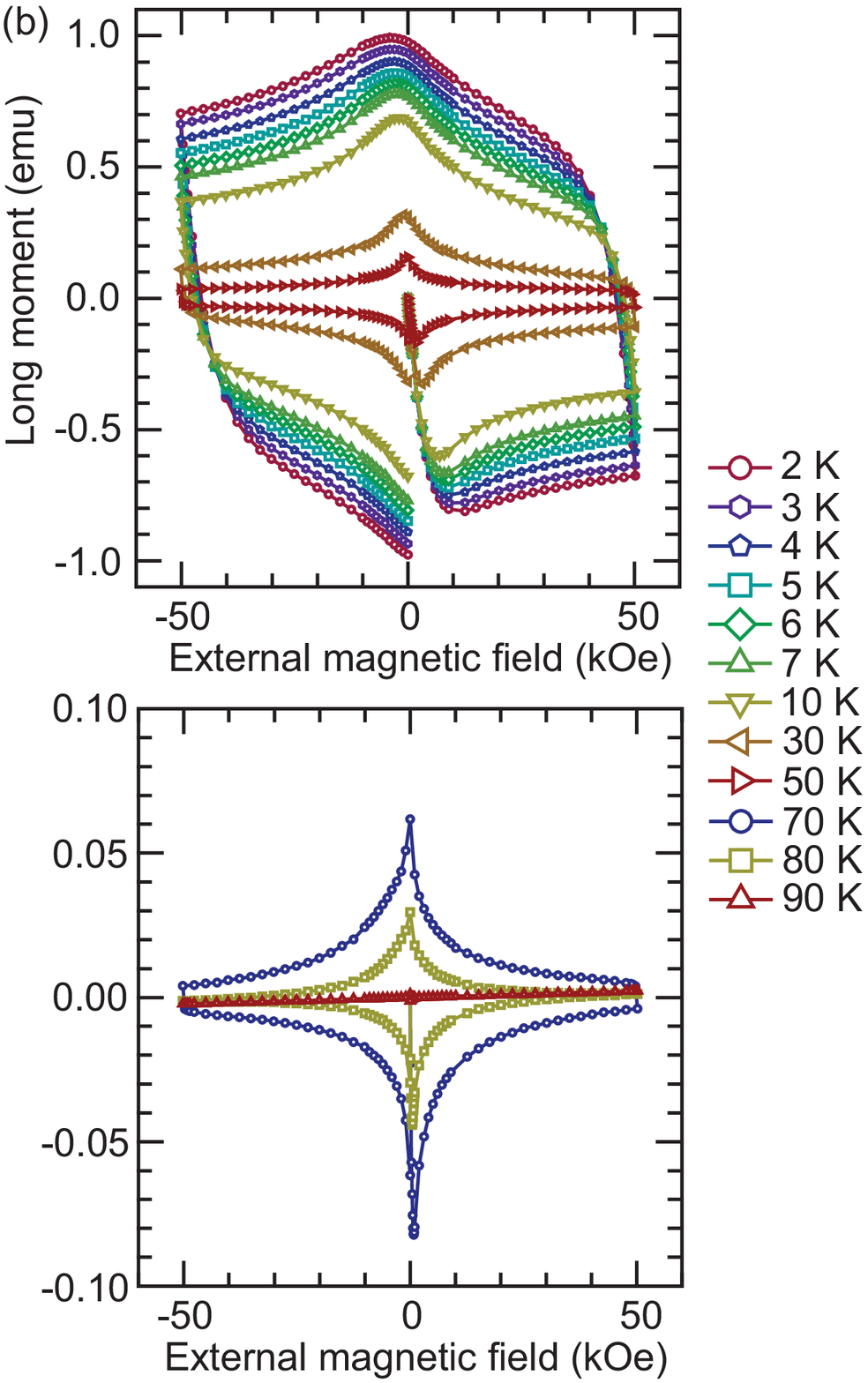}
\caption{Magnetic moment dependence on temperature and applied magnetic field for (a) Nb, (b) YBCO.}
\label{fig:MH_meas}
\end{center}
\end{figure}

\subsection{Critical current density $J_{\mathrm{c}} (T,B)$ estimation}

The critical current density $J_{\mathrm{c}}$ of a superconductor is the maximum electrical current density that can flow through it without causing energy dissipation due to vortex motion. 
The temperature and external magnetic field dependent critical current density can be calculated from the magnetic moment using the Bean model~\cite{Bean-model, Bean-Rev}.
This is a phenomenological model that assumes a spatially uniform critical current density independent of the magnetic field applied. 
Here the current density $j = J_{\mathrm{c}}$ inside the superconductor follows the Amp\'ere's law,
\begin{gather}
\frac{d B_z}{dx} = \mu_0 j, 
\tag{B1} 
\end{gather}
where $z$ is the direction of the magnetic field $B$. 
The image field created inside the superconductor by this current is
\begin{gather}
B_z (x) = \mu_0 J_{\mathrm{c}} x + const, 
\tag{B2} 
\end{gather}
which decreases linearly with the distance from the superconductor surface. 

\begin{figure}[!t]
\begin{center}
\includegraphics[height=5cm, clip]{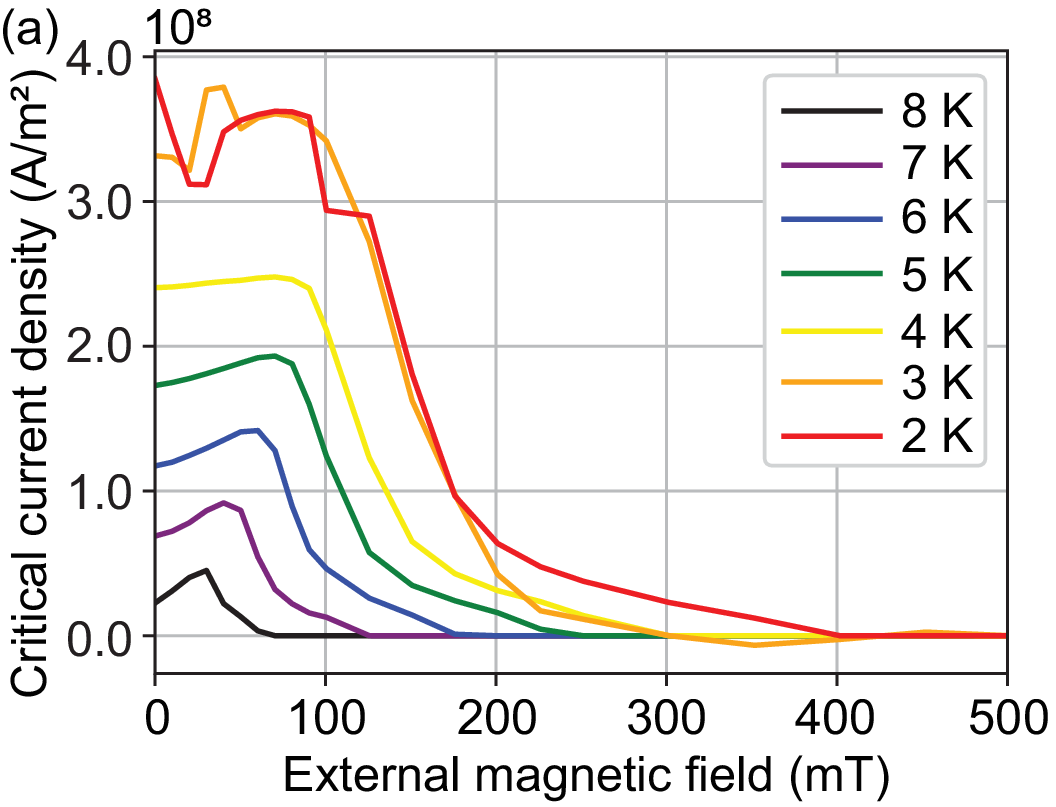}
\vspace{3mm}
\includegraphics[height=5cm, clip]{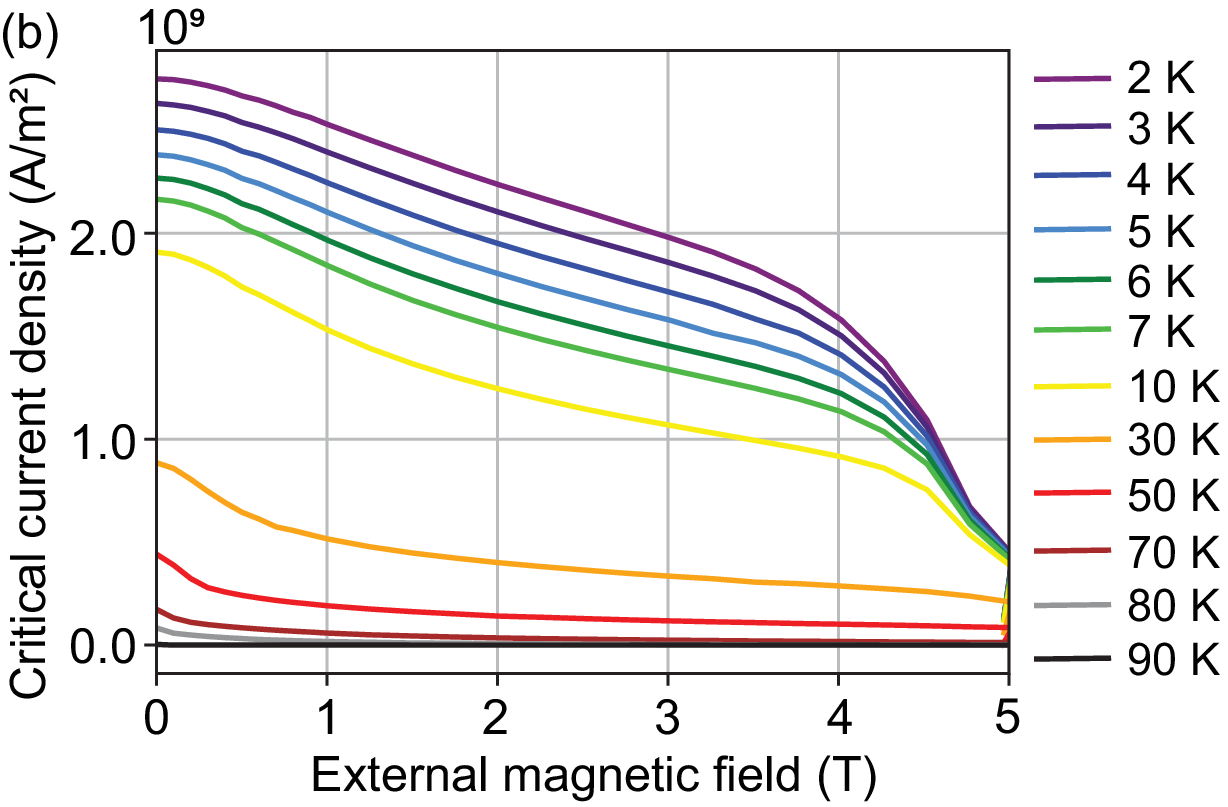}
\caption{Temperature and magnetic field dependence of $ J_{\mathrm{c}} $ for (a) Nb, (b) YBCO.} 
\label{fig:Jc_meas}
\end{center}
\end{figure}

The \textit{M-H} curve of the Nb sample used for $J_{\mathrm{c}}$ measurement is given in Fig.~\ref{fig:MH_meas} (a). 
The fluctuations in the magnetic moment at 2 K and 3 K is due to flux jumps caused by magnetic flux avalanches, the rapid invasion of a large number of quantized magnetic flux due to thermo-magnetic instability.
The negative value of the magnetic moment in the field-decreasing branch indicates that the irreversible magnetic moment is smaller than the reversible one due to weak flux pinning. 
This Nb sample is a cylinder, which creates a cone-shaped magnetic field profile and the magnetic moment is given by
\begin{gather}
 m_{\mathrm{sc}}
= \frac{1}{2} \int r \times J_{\mathrm{c}} \, d r
= J_{\mathrm{c}} \int dr \int dz \, \pi r^2
= \pi J_{\mathrm{c}} \frac{a^3}{3} h_{\mathrm{sc}},  
\tag{B3} 
\end{gather}
where $a$ and $h_{\mathrm{sc}}$ are the radius and height. 
The superconductor magnetization is 
\begin{gather}
M_{\mathrm{sc}} 
= \frac{m_{\mathrm{sc}}}{V} 
= \pi J_{\mathrm{c}} \frac{a^3 h_{\mathrm{sc}} }{3} \, \frac{1}{\pi a^2 h_{\mathrm{sc}}}
= \frac{J_{\mathrm{c}} a}{3}, 
\tag{B4} 
\end{gather}
where $V$ 
the volume of the cylinder. 
Solving this for $J_{\mathrm{c}}$ gives
\begin{gather}
J_{\mathrm{c}} = \frac{3 M_{\mathrm{sc}} }{a}. 
\tag{B5} 
\end{gather}
This critical current density is called the ``magnetic critical current density", which gives the maximum current density that a superconductor can carry before it starts to dissipate energy. 
From Fig.~\ref{fig:MH_meas} (a), 
$m_{\mathrm{sc}} (4 \, \mathrm{K}, \, 0 \, \mathrm{mT}) 
\sim 0.16 \, \mathrm{emu} = 1.6 \times 10^{-4} \, \mathrm{A/m^2}$ 
for a Nb sample of 
$a = 0.835 \, \mathrm{mm}, \, V = 2.28 \times 10^{-9} \, \mathrm{m^3}$
gives 
$ M_{\mathrm{sc}} (4 \, \mathrm{K}, \, 0 \, \mathrm{mT}) \sim 7.0 \times 10^4 \, \mathrm{A/m^2} $ and
$ J_{\mathrm{c}} (4 \, \mathrm{K}, \, 0 \, \mathrm{mT}) \sim 2.5 \times 10^8 \, \mathrm{A/m^2} $.
This is much lower than the value measured for Nb thin films at $ 2.0 \times 10^{10} \, \mathrm{A/m^2} $ at 4.2 K~\cite{Nb-Jc}. 
The temperature and external magnetic field dependence of $J_{\mathrm{c}}$ is given in Fig.~\ref{fig:Jc_meas} (a). 

The \textit{M-H} curve for the YBCO sample used for $J_{\mathrm{c}}$ measurement is given in Fig.~\ref{fig:MH_meas} (b). 
Due to the strong flux pinning force of YBCO, it retains a large amount of trapped magnetic flux after the external field is removed, and the \textit{M-H} curve exhibits a large magnetic hysteresis loop. 
This YBCO sample is a cylindrical structure with a hole and two slits along the radial direction. 
Thus we model our superconductor as two rectangles with width $a$, length $b \, (a<b)$ and height $c$ of which $J_{\mathrm{c}}$ is given by  
\begin{align}
m_{\mathrm{sc}}'
&= 2 \int \int S(x) J_{\mathrm{c}} \, dx dz \notag \\
&= 2 c J_{\mathrm{c}} \int_0^{\frac{a}{2}} [ 4 x^2 + 2x (b-a) ] \, dx  \notag \\
&= \frac{1}{2} J_{\mathrm{c}} a^2 b c \left( 1 - \frac{a}{3b} \right) 
\tag{B6}   
\end{align}
\begin{align}
M_{\mathrm{sc}}' 
&= \frac{m_{\mathrm{sc}}'}{V'} 
= \frac{1}{2} J_{\mathrm{c}} a^2 b c \left( 1 - \frac{a}{3b} \right) \frac{1}{2 abc} \notag \\
&= \frac{1}{4} J_{\mathrm{c}} a \left( 1 - \frac{a}{3b} \right) 
\tag{B7} 
\end{align} 
\begin{align}
J_{\mathrm{c}} = \frac{4 M_{\mathrm{sc}}' }{a \left( 1 - \frac{a}{3b} \right)}
\tag{B8} 
\end{align}
with $S(x)$ being the cross sectional area normal to the external magnetic field, 
$a = ( d_{\mathrm{sc} \, \mathrm{out}} - d_{\mathrm{sc}} )/2, \, 
b = \pi ( d_{\mathrm{sc} \, \mathrm{out}} + d_{\mathrm{sc}} )/4 $. 
From Fig.~\ref{fig:MH_meas} (b), 
$m_{\mathrm{sc}} (4 \, \mathrm{K}, \, 0 \, \mathrm{mT}) 
\sim 0.89 \, \mathrm{emu} = 8.9 \times 10^{-4} \, \mathrm{A/m^2}$ 
for a YBCO sample of 
$a = 0.596 \, \mathrm{mm}, \, b = 2.124 \, \mathrm{mm}, \, V' = 2.63 \times 10^{-9} \, \mathrm{m^3} $
gives
$ M_{\mathrm{sc}} (4 \, \mathrm{K}, \, 0 \, \mathrm{mT}) \sim 3.4 \times 10^5 \, \mathrm{A/m^2} $ and
$ J_{\mathrm{c}} (4 \, \mathrm{K}, \, 0 \, \mathrm{mT}) \sim 2.5 \times 10^9 \, \mathrm{A/m^2} $. 
This is higher than the critical current density of Nb, and much lower than the critical current density measured for YBCO wires at $ 2.0 \times 10^{11} \, \mathrm{A/m^2} $ at 4.2 K~\cite{YBCO-Jc}. 
The temperature and external magnetic field dependence is given in Fig.~\ref{fig:Jc_meas} (b). 
To measure the full phase diagram of $J_{\mathrm{c}}$, it is necessary to measure the full magnetic moment hysteresis curve using high enough external magnetic fields. 
Since the external magnetic field of 5 T was not high enough, artifacts are introduced in the form of abrupt decrease in $J_{\mathrm{c}}$ above 4 T for temperatures below 10 K. 

\begin{figure}[!t]
\begin{center}
\includegraphics[height=5cm, clip]{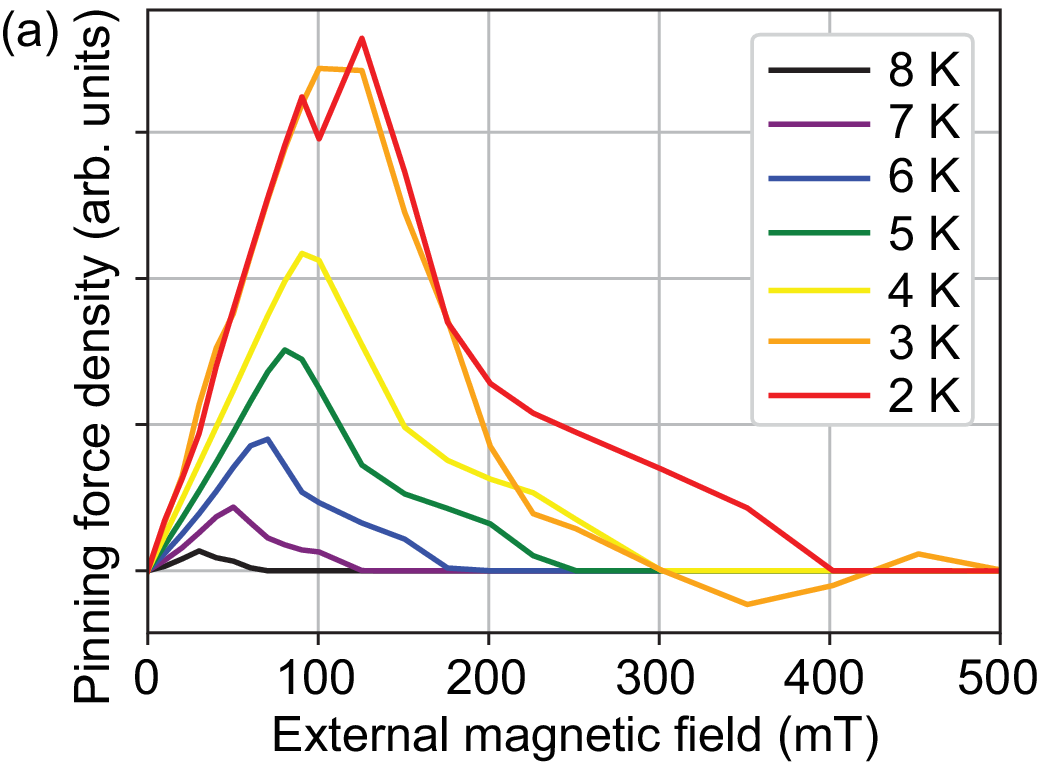}
\vspace{3mm}
\includegraphics[height=5cm, clip]{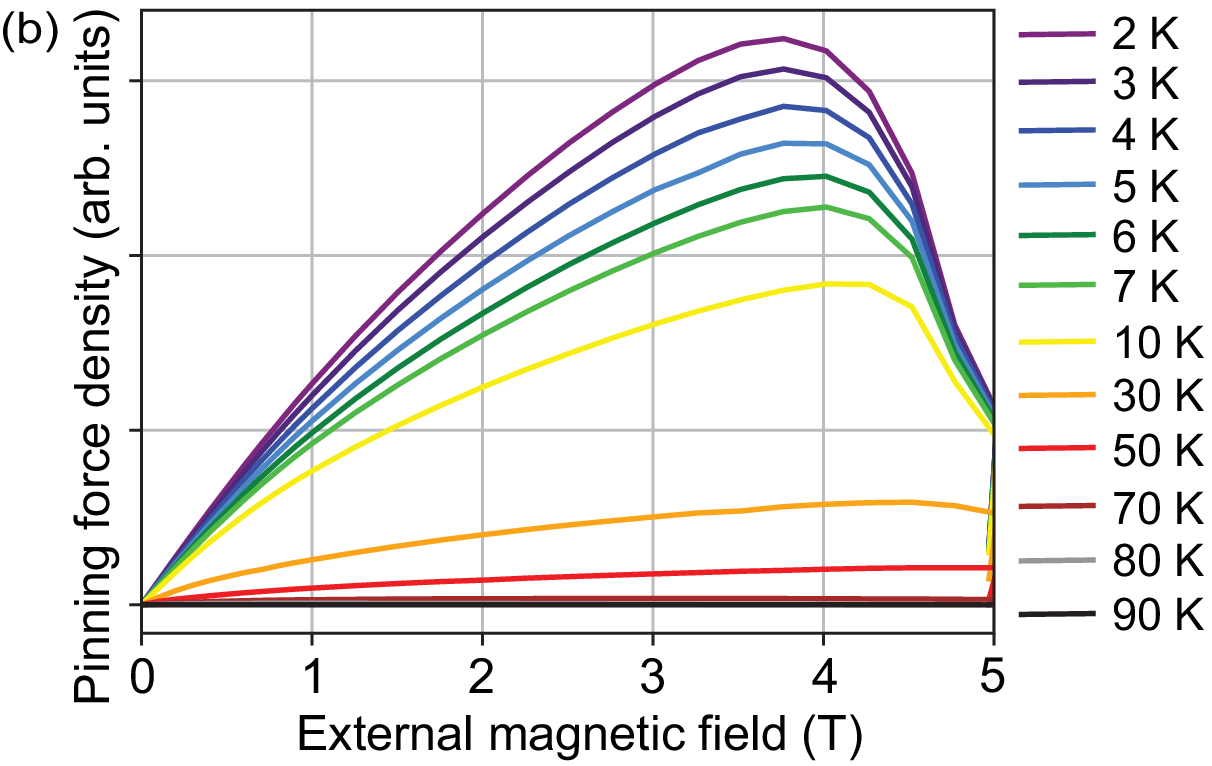}
\caption{Temperature and magnetic field dependence of $F_p$ for (a) Nb, (b) YBCO.} 
\label{fig:Fp_meas}
\end{center}
\end{figure}

\subsection{Pinning force density $ F_p (T, B)$ estimation}

When a magnetic field is applied to a type II superconductor, vortices, which each carry a quantized magnetic flux, can penetrate the material. 
For low magnetic fields, the superconductor is in the vortex solid state, where vortices in a superconductor form a lattice. 
In the absence of any external forces, these vortices are pinned by defects and impurities in the material, resulting in a static arrangement. 
The superconductor is in a mixed state where both superconducting and normal conducting regions coexist within the material. 
When the temperature or magnetic field exceeds the critical limit, the superconductor transitions into a vortex liquid state.
Vortices move freely throughout the superconductor in a disordered manner, due to thermal fluctuations, external magnetic fields, or interactions with each other, exhibiting fluid-like behavior. 
The motion of vortices causes substantial energy dissipation, and this phase is therefore not suitable for levitation.

This critical magnetic field can be estimated by the flux pinning force density $ F_p = J_{\mathrm{c}} B $, which is a measure of the maximum pinning force per unit volume that can be exerted on vortices in a superconductor before they start to move (Fig.~\ref{fig:Fp_meas}). 
This flux pinning force increases with applied magnetic field, because the magnetic field increases the number of vortices, enhances the interaction between the vortices and the pinning centers, making it more difficult for them to move or escape their pinned positions. 
However, beyond a certain magnetic field strength $H_{\mathrm{fp}}$, the stronger pinning centers become fully occupied or saturated, and additional vortices cannot be effectively pinned, reducing the ability to trap and immobilize vortices. 
Fig.~\ref{fig:Fp_meas} (a) shows that magnetic fields above $H_{\mathrm{fp}} \sim 100$ mT may cause vortex movement for temperatures below 4 K. 
This critical magnetic field decreases with increasing temperature. 
For $B_z \leq 75$ mT used for YIG levitation, vortex movement induced dissipation will arise for $T > 6$ K.  
Fig.~\ref{fig:Fp_meas} (b) together with Fig.~\ref{fig:MH_meas} (b) and Fig.~\ref{fig:Jc_meas} (b) shows that the peak in pinning force density is due to the artifacts in $J_{\mathrm{c}}$ measurement introduced by insufficient maximum external magnetic field strength. 
Thus $H_{\mathrm{fp}} > 5$ T, and there is no vortex movement for magnetic fields below 5 T.

\begin{figure}[!b]
\begin{center}
\includegraphics[height=5cm, clip]{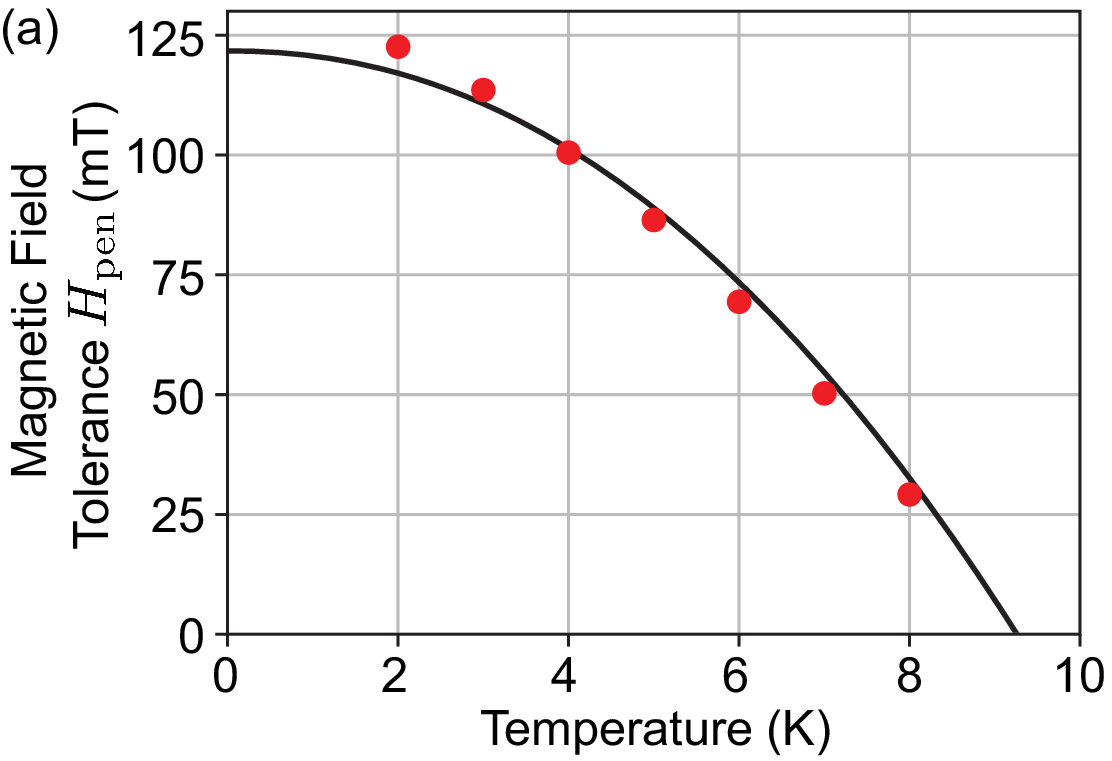}
\vspace{3mm}
\includegraphics[height=5cm, clip]{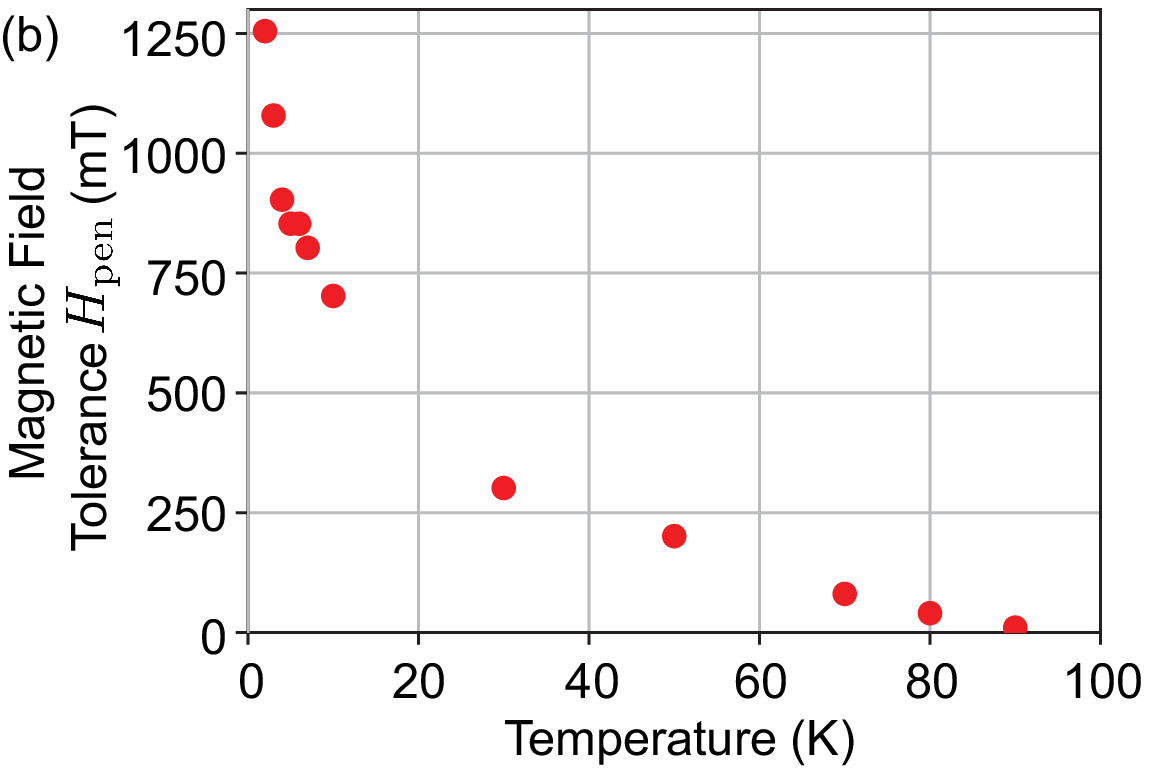}
\caption{Temperature dependence of $ H_{\mathrm{pen}} $ for (a) Nb, (b) YBCO.} 
\label{fig:Hpen_meas}
\end{center}
\end{figure}

\subsection{Magnetic field tolerance $ H_{\mathrm{pen}} (T)$ estimation}

We estimate $ H_{\mathrm{pen}} $, the maximum magnetic field our superconductor sample could expel by perfect cancellation of their magnetic moments, which depends not only on the type of superconductor, purity, and structural properties but also shape. 
This can be estimated from Fig.~\ref{fig:MH_meas} as the external magnetic field when the modulus of the magnetic moment of a superconductor is maximum (Fig.~\ref{fig:Hpen_meas}). 

For pure Nb, $ H_{\mathrm{pen}} $ follows the same temperature dependence curve as the critical magnetic field $H_{\mathrm{c}}$, and the experimental data can be fitted to 
\begin{gather}
H_{\mathrm{pen}} (T) = H_{\mathrm{pen}} (0) \left( 1 - \left( \frac{T}{T_{\mathrm{c}}} \right)^\gamma \right) 
\tag{B9}
\end{gather}
with $ T_{\mathrm{c}} = 9.26$ K, $\gamma = 2.13 $ predicted from the BCS (Bardeen–Cooper–Schrieffer) theory. 
This shows that the Nb is in a diamagnetic state for magnetic fields below $ H_{\mathrm{pen}} $. 
The maximum magnetic field our sample can tolerate at 0 K is $H_{\mathrm{pen}} (0) = 122 \pm 21 $ mT.
Assuming the temperature of the Nb bulk is equal to the base plate, for YIG levitation conditions of $T \leq 4.6$ K, $H_{\mathrm{pen}} (4.6 \, \mathrm{K}) \sim 94 $ mT is larger than the magnetic fields applied. 
However, since the external magnetic field $B_z$ is close to $H_{\mathrm{pen}}$, local damage in our Nb sample can lead to flux penetrations and increase effective Nb hole inner diameter $ d_{\mathrm{sc}} $ as expected from trap frequency measurements. 

The magnetic field tolerance $ H_{\mathrm{pen}} (4.6 \, \mathrm{K}) \sim 900 $ mT proves the YBCO sample expelled magnetic flux completely during the entire experiment (Fig.~\ref{fig:Hpen_meas} (b)). 
The sharp increase of $H_{\mathrm{pen}}$ at low temperatures is due to strong flux pinning in YBCO.
Further research on how strongly pinned vortices affect the eddy current damping at higher external magnetic fields will be conducted elsewhere.

\section{Appendix C: Three-dimensional finite element method simulations}

A three-dimensional COMSOL model was constructed using the parameters in Table~\ref{tbl_exp_param} to simulate the dynamics of the levitated YIG sphere. 

\begin{figure}[!t]
\begin{center}
\includegraphics[height=5cm, clip]{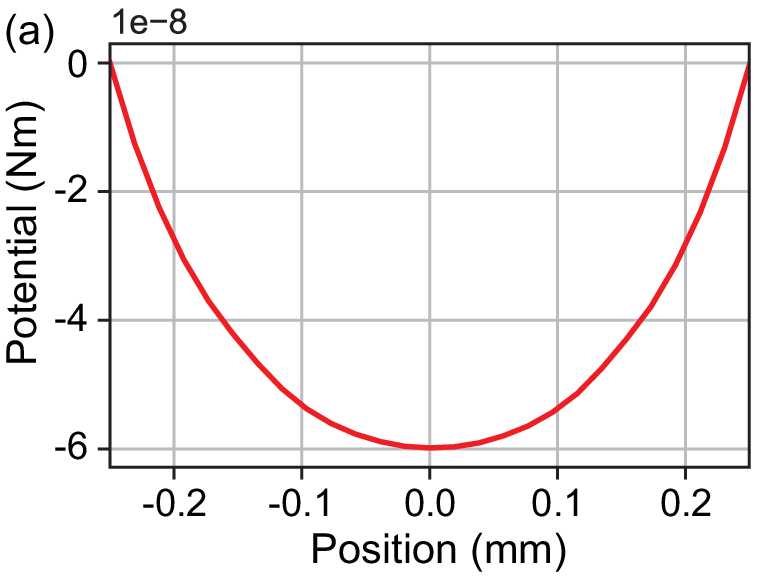}
\vspace{3mm}
\includegraphics[height=5cm, clip]{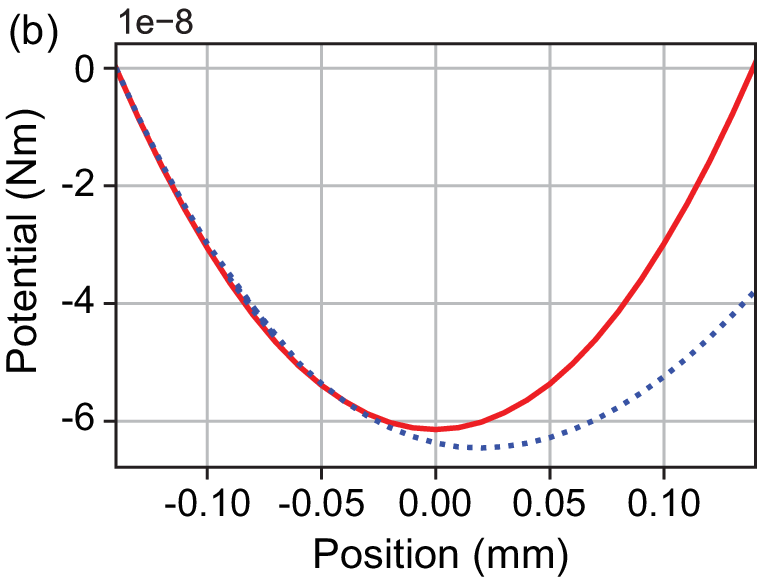}
\caption{Trap potentail of $d = 0.5$ mm YIG sphere under an external magnetic field of $B_z = 37.5$ mT 
(a) vertical trapping potential ($z$ direction in Fig. 1 of main text), 
(b) horizontal trapping potential in red solid line: perpendicular to the slit ($y$ direction in Fig. 1 of main text), blue dotted line: in the direction of the slit ($x$ direction in Fig. 1 of main text).} 
\label{fig:potential}
\end{center}
\end{figure}

\subsection{Trap potential}

First the three-dimensional magnetic field distribution is calculated by a stationary study, using the Amp\'ere's law.
Since the external magnetic field applied is lower than the saturation magnetization of pure YIG at low temperatures $M_s$ = 196 kA/m ($\mu_0 M_s \sim 246$ mT), we use the relative permeability $ \mu_\gamma = 32 $~\cite{YIGmu-init} to calculate the constitutive relation $ B = \mu_0 \mu_\gamma H $ used during this calculation. 
This allows the YIG magnetization to both spatially and temporary respond to the changing magnetic field, which is consistent with the fact that it cannot be treated as a single domain magnetic dipole. 
The resulting force on the YIG sphere was calculated by numerically integrating the Maxwell stress tensor
\begin{gather}
\mathbf{F}_{em} = \int \int \frac{\mathbf{B}_n \cdot \mathbf{B}_n }{2 \mu_0} \, dS 
\tag{C1}
\end{gather}
where $\mathbf{B}_n $ is the normal component of the magnetic flux density. 

In order to accurately calculate the force on the YIG sphere, constructing a fine enough mesh is crucial.
We calculated the forces on the YIG sphere with varying maximum and minimum mesh sizes to resolve the small regions between the sphere and superconducting wall.
A mesh conversion analysis showed that a minimum mesh size of $d / 100 = 5 \, \mu$m and largest mesh size of $d / 20 = 25 \, \mu$m was required to calculate the force on a YIG sphere of $d = 0.5$ mm with an accuracy of over two scientific digits. 
This was used to calculate the force on the YIG sphere as a function of its position.
Subsequently, this was numerically integrated to obtain the trapping potential (Fig.~\ref{fig:potential}). 
This shows that the YIG sphere is indeed trapped in a harmonic potential with its center shifted around 20 $\mu$m from the center of the hole towards the slit.

\subsection{Trap frequency}

The dynamics of the YIG sphere is simulated by a time domain study that calculates first-order ordinary differential equations 
\begin{gather}
\dot{v} = F_{em} / m \tag{C2} \\
\dot{u} = u \tag{C3}
\end{gather}
where $m$ is the mass, $u$ the position, $v$ the velocity of the YIG sphere respectively. 
The resulting oscillations are fitted to a sine-curve to determine the trap frequency.

\subsection{Eddy current estimation}

Eddy current damping occurs when a magnetic object, in this case the YIG sphere, moves relative to a conductor, in this case the bobbin and surrounding oxygen free copper jigs. 
A damping force that opposes the motion of the YIG sphere arises from eddy currents, which create their own magnetic field that opposes the original magnetic field. 
The magnitude of this damping force is proportional to the conductivity of the conductor, strength of the magnetic field, and velocity of the magnet. 
It has been the dominant loss in many magneto-mechanical systems. 

A time-dependent solver in COMSOL is used to calculate the electric energy dissipated per oscillation in the YIG sphere, superconductor coil bobbin, oxygen free copper lid, and base plate. 
The magnetic vector potential is calculated from
\begin{gather}
( j \omega \sigma - \omega^2 \varepsilon ) \mathbf{A} + \nabla \left( \frac{1}{\mu} \nabla \times \mathbf{A} \right) = 0 
\tag{C4}
\end{gather}
where $\sigma$ is the conductivity, $\varepsilon$ the permittivity, $\mu$ the permeability, $\omega$ the trap angular frequency, and $ \delta = \sqrt{ 2/ \omega \mu \sigma }$ the skin depth. 
For an oxygen free copper plate at low temperatures, the skin depth is nearly equal to the YIG diameter $ \delta \sim d $. 
For the YIG sphere which is an insulator, the skin depth $\delta > 1$ km is larger enough than the objects. 
In these cases, the eddy current can be calculated by volume integration of the Poynting vector as 
\begin{gather}
P_{\mathrm{eddy}} = \frac{1}{2} ( \mathbf{J}_S \cdot \mathbf{E}* ) 
\tag{C5}
\end{gather}
where $ \mathbf{J}_S $ is the induced current and $\mathbf{E}$ is the electric field inside the object. 

\begin{figure}[t]
\begin{center}
\includegraphics[height=4cm, clip]{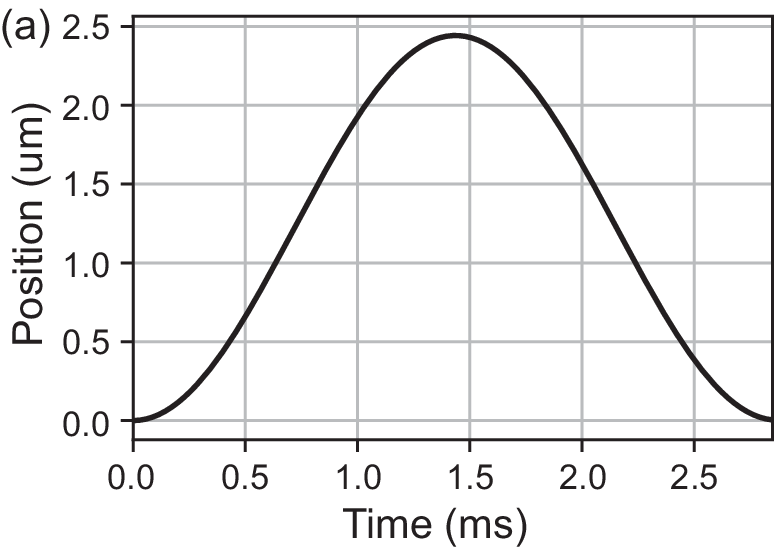}
\vspace{3mm}
\includegraphics[height=4cm, clip]{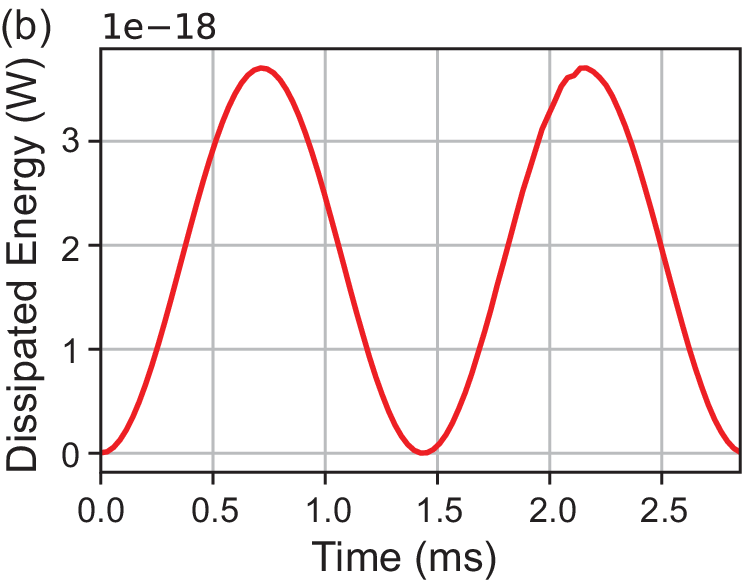}
\vspace{3mm}
\includegraphics[height=4cm, clip]{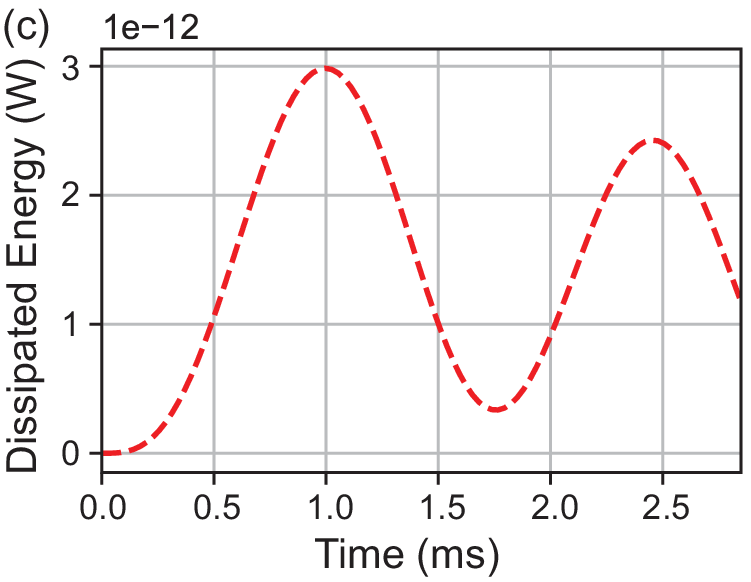}
\caption{Example of time dependent electro-magnetic energy loss during a single oscillation for $B_z = 62.5$ mT. 
(a) YIG position, 
(b) energy dissipation in YIG sphere, 
(c) total energy dissipation in superconductor coil bobbin, base plate and oxygen free copper lid.} 
\label{fig:eddy_current_est}
\end{center}
\end{figure}

A time-dependent solver in COMSOL is used to calculate the electrical losses in both the YIG sphere, bobbin, lid and base plate for a single oscillation (Fig.~\ref{fig:eddy_current_est}). 
The YIG oscillation causes eddy currents in both the YIG sphere and its surroundings, which causes a displacement dependent dissipation. 
Since this dissipation in the bobbin and surrounding jigs is a first order induction effect caused by the YIG motion, while the eddy current in the YIG sphere is a second order induction effect caused by the magnetic field change due to eddy current in the bobbin and surrounding jigs, the former is larger than the latter. 
In the time dependent study, an oscillation is divided into time slots of $ \delta t = 1/ f / 100 $ s, and the energy dissipation during each temporal duration is calculated. 
The energy dissipation per oscillation $ \Delta E_{\mathrm{eddy}} $ is given by the average energy loss within the oscillation. 
Finally the $Q$-factor $ Q_{\mathrm{eddy}} = 2 \pi E_t / (\Delta E_{\mathrm{eddy}}) $ is estimated by comparing this to the kinetic energy $ E_t = m A_z^2 \omega_z^2 / 2 $ where $m = \rho_y \, 4/3 \pi a^3$ is the mass of the YIG sphere and $A_z$ is the amplitude of oscillation taken from the position of the YIG.

\end{document}